\documentstyle[aasms4,psfig]{article}

\begin{document}

\title{The X-Ray Spectrum and Global Structure \\
        of the Stellar Wind in Vela X-1}
\author{Masao Sako\altaffilmark{1}, Duane A. Liedahl\altaffilmark{2},
        Steven M. Kahn\altaffilmark{1}, and Frits Paerels\altaffilmark{1,3}}

\altaffiltext{1}{Columbia Astrophysics Laboratory and Department of Physics,
                 Columbia University, 538 West 120th Street, New York, NY
                 10027; masao@astro.columbia.edu (MS), \\
                 skahn@astro.columbia.edu (SMK)}
\altaffiltext{2}{Department of Physics and Space Technology,
                 Lawrence Livermore National Laboratory,
                 P.O. Box 808, L-41, Livermore,  CA  94550;
                 duane@serpens.llnl.gov}
\altaffiltext{3}{SRON Laboratory for Space Research, Sorbonnelaan 2,
                 3584CA, Utrecht, The Netherlands; F.Paerels@sron.nl}

\begin{abstract}

  We present a quantitative analysis of the X-ray spectrum of the eclipsing
  high mass X-ray binary Vela X-1 (4U 0900 -- 40) using archival data from the
  {\small {\it ASCA}} Solid-State Imaging Spectrometer. The observation covers
  a time interval centered on eclipse of the X-ray pulsar by the companion.
  The spectrum exhibits two distinct sets of discrete features: (1)
  recombination lines and radiative recombination continua from mostly
  hydrogenic and helium-like species produced by photoionization in an
  extended stellar wind; and (2) fluorescent K-shell lines associated with
  near-neutral species also present in the circumsource medium.  These
  features are superposed on a faint continuum, which is most likely
  nonthermal emission from the accreting neutron star that is scattered into
  our line of sight by free electrons in the wind.

  Using a detailed spectral model that explicitly accounts for the
  recombination cascade kinetics for each of the constituent charge states, we
  are able to obtain a statistically acceptable ($\chi_{r}^{2}=0.88$) fit to
  the observed spectrum and to derive emission measures associated with the
  individual K-shell ions of several elements. From calculations of the
  ionization balance using the photoionization code, {\small XSTAR}
  \markcite{kall95}(Kallman \& Krolik 1995), we assign ionization parameters,
  $\xi$, to several ions, and construct a differential emission measure
  ({\small DEM}) distribution. The {\small DEM} distribution spans a broad
  range in $\xi$ ($\Delta\log\xi \gtrsim 2$), and is centered around $\log\xi
  = 2.5$. We find that the total emission measure of the visible portion of
  the highly ionized wind is $\sim3 \times 10^{56} ~\rm{cm^{-3}}$.

  The qualitative aspects of the inferred {\small DEM} distribution are
  consistent with a wind model derived from the Hatchett \& McCray
  \markcite{hatc77}(1977) picture of an X-ray source immersed in a stellar
  wind with a generalized Castor, Abbott, \& Klein \markcite{cast75}(1975)
  velocity profile. Using this formalism, theoretical {\small DEM}
  distributions, parameterized only by a mass loss rate and a wind velocity
  profile, are calculated and used to predict the detailed X-ray spectrum,
  which is then compared to the {\small {\it ASCA}} data.  Again, we find a
  statistically acceptable fit ($\chi_{r}^{2} = 1.01$), with a best-fit mass
  loss rate of $\sim 2.7 \times 10^{-7} ~M_\odot ~\rm{yr^{-1}}$.  This is
  approximately a factor of 10 lower than previous estimates of the mass loss
  rate for the Vela X-1 companion star, which have primarily been determined
  from \ion{C}{4} and \ion{Si}{4} P~Cygni profiles, and X-ray absorption
  measurements.  We argue that this discrepancy can be reconciled if the X-ray
  irradiated portion of the wind in Vela X-1 is structurally inhomogeneous,
  consisting of hundreds of cool, dense clumps embedded in a hotter, more
  ionized gas.  Most of the mass is contained in the clumps, while most of the
  wind volume ($> 95\%$) is occupied by the highly ionized component.  We show
  quantitatively, that this interpretation is also consistent with the
  presence of the X-ray fluorescent lines in the {\small {\it ASCA}} spectrum,
  which are produced in the cooler, clumped component.

\end{abstract}


\section{Introduction}

  In wind-driven high mass X-ray binaries ({\small HMXB}s), an intense X-ray
  continuum flux is produced through gravitational capture of a stellar wind
  by a neutron star or a black hole. A small fraction ($\sim 0.1\%$) of the
  total mass lost by the companion is accreted onto the compact object, and a
  fraction of the gravitational potential energy is then converted into
  X-radiation, which ionizes and heats the surrounding gas.  The wind
  reprocesses hard X-rays from the compact object, resulting in discrete
  emission lines and continuum radiation.  X-ray emission lines can be
  produced in a wide range of ionization states, ranging from cold fluorescent
  emission line regions to highly ionized recombination line regions.  The
  resulting spectrum carries information that should, in principle, allow us
  to derive physical parameters that characterize the state and structure of
  the circumsource material.

  Vela X-1, the archetypal wind-fed X-ray pulsar, has been the most
  well-studied object in its class since its discovery
  \markcite{chod67}(Chodil et al.\ 1967).  It is an eclipsing {\small HMXB}
  with a pulse period of 283 s \markcite{mccl76}(McClintock et al.\ 1976) and
  an orbital period of 8.96 days \markcite{form73}(Forman et al.\ 1973).  The
  optical counterpart, {\small HD} 77581, is a B0.5Ib supergiant
  \markcite{bruc72,hilt72}(Brucato \& Kristian 1972; Hilter, Werner, \& Osmer
  1972) with an inferred mass loss rate lying in the range (1 -- 7) $\times
  10^{-6} ~M_{\odot} ~\rm{yr^{-1}}$
  \markcite{hutc76,dupr80,kall82,sada85,sato86} (Hutchings 1976; Dupree et
  al.\ 1980; Kallman \& White 1982; Sadakane et al.\ 1985; Sato et al.\ 1986).
  Its intrinsic X-ray luminosity is $\sim 10^{36} ~\rm{erg~s^{-1}}$, and is
  consistent with accretion from laminar flow past a stellar object
  \markcite{bond44}(Bondi \& Hoyle 1944) for a mass loss rate and velocity
  structure characterizing a typical {\small OB} supergiant.

  Prior to the launch of {\small {\it ASCA}}, X-ray emission line studies of
  Vela X-1 were typically limited to the iron K complex near 6.4 keV; emission
  lines from lighter elements were not resolved
  \markcite{whit83,ohas84,lewi92}(White, Swank, \& Holt 1983; Ohashi et al.\
  1984; Lewis et al.\ 1992).  The solid state detectors onboard {\small {\it
  ASCA} }are capable of resolving bright spectral features from H-like and
  He-like ions, although, for the most part, distinguishing the effects
  produced by various emission mechanisms is still problematic.  The first
  analysis of the {\small {\it ASCA}} spectrum of Vela X-1, which emphasized
  line emission, was presented by Nagase et al.\ \markcite{naga94}(1994,
  hereafter N94). The hard X-ray spectrum (3 -- 10 keV) is highly absorbed and
  contains bright iron K fluorescent lines. During eclipse ($-0.1 \lesssim
  \phi \lesssim \rm{+}0.1$), the continuum radiation from the compact source
  is occulted by the companion, and emission lines that form in extended
  regions gain contrast with respect to the ionizing continuum, resulting in
  the line-dominated appearance of the soft X-ray band (0.5 -- 3 keV). N94
  tentatively identified most of the bright emission lines, while some of the
  fainter mid-$Z$ fluorescent lines were identified in a later article
  \markcite{naga97}(Nagase 1997).

  Although it was suggested in N94 that the more highly ionized X-ray emission
  lines observed in the Vela X-1 spectrum are formed by cascades following
  radiative recombination, those authors did not have access to an
  appropriate, self-consistent, recombination spectral model suitable for a
  detailed quantitative analysis of the emission line intensities.  While it
  is all but certain that many X-ray lines in {\small HMXB}s are recombination
  lines, it is, in fact, difficult to distinguish recombination from other
  emission mechanisms in {\small CCD} spectra, thereby disallowing a
  quantitative interpretation of the spectra.  \markcite{lied96}Liedahl \&
  Paerels \markcite{lied96}(1996, hereafter LP96) applied a suitable
  recombination cascade model to the interpretation of the {\small {\it ASCA}}
  spectrum of the {\small HMXB}, Cyg X-3. Infrared line spectroscopy of this
  object indicates the presence of a dense stellar wind from a Wolf-Rayet
  companion \markcite{vank95}(van Kerkwijk et al.\ 1992).  As demonstrated in
  LP96, the characteristics of the X-ray line spectrum, including the shape
  and magnitude of the emission measure distribution conform to what one would
  expect according to a simple model of a photoionized stellar wind in an
  X-ray binary system.

  The purpose of this paper is to extend and refine the analyses of LP96 and
  N94, and to demonstrate that the X-ray spectrum of Vela X-1 can be
  interpreted in terms of an explicit stellar wind model.  Our approach
  improves upon the method of LP96 in several ways.  First we take explicit
  account of wind acceleration according to a modified Castor, Abbott, \&
  Klein \markcite{cast75}(1975, hereafter {\small CAK}) wind model.  We model
  the orbital phase dependence of the apparent differential emission measure,
  taking advantage of the fact that the system parameters (binary separation,
  companion radius, inclination angle, ephemeris, etc.) are much better
  constrained in Vela X-1 than in Cyg X-3.  The LP96 method of fitting the
  data with model ionic spectra, each of which is assumed to be at a single
  temperature, is adopted as an initial step in the course of our
  analysis. But we also analyze the data with a more elaborate model that
  accounts for temperature variations in each ionization zone.  Using the
  Hatchett \& McCray \markcite{hatc77}(1977, hereafter HM77) picture of an
  X-ray source immersed in a stellar wind, coupled with a generalized {\small
  CAK} wind, we show that the shape and magnitude of the differential emission
  measure ({\small DEM}) distribution in {\small HMXB}s depend sensitively on
  stellar wind parameters. From a grid of model {\small DEM} distributions,
  theoretical spectra are generated and compared directly to the spectral
  data, leading to the determination of the rate at which highly ionized
  material is accelerated and the corresponding mass loss rate.

  This paper is organized as follows.  In \S2, we describe the data reduction
  performed, the details of our explicit spectral model, and the derivation of
  the empirical {\small DEM} distribution for the ionized stellar wind.  In
  \S3, we discuss physical models for the stellar wind, and our derivation of
  wind parameters.  In \S4, we present a quantitative analysis of the K-shell
  fluorescent lines.  Finally, in \S5, we discuss the implications of our
  results in terms of previous models for the Vela X-1 system, and show that
  all of the data can be reconciled through a model of an inhomogeneous wind
  were most of the mass is contained in relatively dense cool clumps, while
  most of the volume is occupied by a lower density, more highly ionized
  tenuous medium.

\section{Spectral Analysis}

\subsection{Data Reduction}

  Vela X-1 was observed with {\small {\it ASCA}} on three occasions: twice in
  1993 and once in 1995.  The first set of observations in 1993 was performed
  during the performance verification phase, and covered the entire transition
  from ingress to egress ($-0.13 \leq \phi \leq \rm{+}0.16$).  The observation
  in 1995 covered only part of the eclipse phase ($-0.05 \leq \phi \leq
  \rm{+}0.10$).  For the purpose of our analysis, we only use data obtained in
  1993.  We separate the observations into three phase ranges: pre-eclipse
  ($-0.13 \leq \phi \leq -0.10$), eclipse ($-0.10 \leq \phi \leq \rm{+}0.10$),
  and post-eclipse ($\rm{+}0.12 \leq \phi \leq \rm{+}0.16$).  For the eclipse
  phase, we use data from {\small SIS}0 and {\small SIS}1.  For the pre- and
  post-eclipse phases, which we analyze principally to derive the intrinsic
  X-ray continuum spectrum of the compact source, we only use data from
  {\small GIS}2.

  The data reduction was performed with {\small FTOOLS} v4.0.  The data were
  screened through standard criteria: rejection of hot flickering pixels and
  data contaminated by bright earth, event selections based on grade, etc.
  The observation during eclipse was performed in 4-{\small CCD} mode with all
  chips collecting a significant number of counts.  In order to maximize
  statistical quality, source and background counts were separately extracted
  from each individual chip. Data, background, response matrix, and ancillary
  response matrix files were generated for each {\small CCD} chip and were
  combined using standard methods.  The resulting spectra contained $1.1
  \times 10^4$ and $9.0 \times 10^3$ counts from {\small SIS}0 and {\small
  SIS}1, respectively, collected during a total exposure time of 38 ksec.  The
  net exposure times of the pre- and post-eclipse phases were 6.0 and 8.4
  ksec, which contained $2.5 \times 10^4$ and $7.4 \times 10^4$ counts,
  respectively.

\subsection{Spectral Model}

  Our spectral model for Vela X-1 is comprised of three principal components:
  (1) continuum radiation from the neutron star, which consists of both direct
  emission from the central source, absorbed through the intervening wind, and
  scattered radiation from free electrons in the extended wind; (2) discrete
  recombination line and radiative recombination continua ({\small RRC}) from
  highly ionized ions; and (3) fluorescent lines from near-neutral ions.  Our
  treatment of each of these components is described in the following
  subsections.

\subsubsection{Continuum Emission}

  The spectra of Vela X-1 require two continuum components.  The intrinsic
  continuum radiation, which originates near the neutron star, is highly
  absorbed as it propagates through the surrounding stellar wind.  A fraction
  of this radiation is scattered into our line of sight through electron
  scattering and is less absorbed relative to that of the direct continuum.
  The direct continuum is also required by the eclipse spectrum, since near
  eclipse ingress ($\phi \sim -0.10$) and egress ($\phi \sim \rm{+}0.10$), the
  line of sight to the compact object is not entirely blocked by the
  companion.  The monochromatic photon flux (photon cm$^{-2}$ s$^{-1}$
  keV$^{-1}$) of the total observed continuum radiation, therefore, has the
  following functional form.
\begin{equation}
  F_{\rm{cont}}(E) = A^{\rm{scat}}e^{-\sigma(E)N^{\rm{scat}}_H}
                     E^{-\Gamma}_{\rm{ke\!V}}
                     \hspace{0.05in} + \hspace{0.05in}
                     A^{\rm{dir}}e^{-\sigma(E)N^{\rm{dir}}_H}
                     E^{-\Gamma}_{\rm{ke\!V}},
\end{equation}
  where $A^{\rm{scat}}$ and $A^{\rm{dir}}$ are the normalizations (in units of
  photon cm$^{-2}$ s$^{-1}$ keV$^{-1}$ at 1 keV) of the scattered and direct
  components, respectively.  Since electron scattering should not appreciably
  alter the spectral shape of the continuum, the photon indices of the two
  components are set equal.  The absorption cross sections are adopted from
  Morrison \& McCammon \markcite{mori83}(1983) and we assume solar abundances
  for the absorbing medium.

\subsubsection{Discrete Recombination Emission}

  Our recombination cascade model is an outgrowth of that described in LP96,
  which was successfully applied to the analysis of the {\small {\it ASCA}}
  spectrum of Cyg X-3.  Several updates have been made, and the current
  version includes line emissitivies of K-shell carbon, nitrogen, oxygen,
  neon, magnesium, silicon, sulfur, argon, calcium, and iron, as well as the
  iron L species (\ion{Fe}{17} -- {\small XXIV}), and the associated {\small
  RRC}.  We have written an interface to the {\small XSPEC} spectral analysis
  package \markcite{arna96}(Arnaud 1996) that calculates line and {\small RRC}
  emissivities for a given ionization stage and electron temperature.  Each
  ion is treated as an individual model component in {\small XSPEC}, with the
  temperature and normalization as adjustable parameters.

  Atomic structure and transition rates are calculated with the atomic physics
  package, {\small HULLAC} (Hebrew University/Lawrence Livermore Atomic Code)
  \markcite{klap77}(Klapisch et al.\ 1977). Ionic spectra are calculated for
  each ion using radiative cascade models that include detailed atomic
  structure through $n=6$ and averaged structure for $n =$ 7 -- 10.  The $n =
  10$ level is a ``superlevel'' that accounts for recombination into shells
  with $n > 10$.  Radiative rates, including {\small E}1, {\small E}2, {\small
  M}1, and {\small M}2 transitions are calculated with {\small HULLAC}.  For
  the 2{\small E}1 two-photon decay rates of hydrogenic $2s_{1/2}$ states, we
  use Parpia \& Johnson \markcite{parp82}(1982).  We assume an $E^{-3}$
  dependence in the photoionization cross sections, where threshold cross
  sections and ionization potentials are adopted from Saloman, Hubble, \&
  Scofield \markcite{salo88}(1988).  The {\small RRC} are calculated from the
  photoionization cross sections using the Milne relation, assuming that the
  recombining electrons are Maxwellian distributed.  The line and {\small RRC}
  emissivities are self-consistently calculated for each ionic species to
  produce the correct line-{\small RRC} ratios.

  The {\it specific line power} $S_{ul}(T)$ for a radiative transition
  connecting an upper level $u$ to a lower level $l$ is given by
  $S_{ul}(T)=\eta_{ul}~\alpha_{RR}(T)$, where $\eta_{ul}$ is the fraction of
  recombinations onto the relevant charge state that produce the transition $u
  \rightarrow l$, and $\alpha_{RR}$ is the total radiative recombination
  ({\small RR}) rate coefficient for that charge state.  The dimensions of
  $S_{ul}$, like $\alpha_{RR}$, are cm$^{3}$ s$^{-1}$.  We calculate $S_{ul}$
  by solving the set of coupled rate equations, assuming that only the ground
  state of the recombining ion is populated.  This is an excellent
  approximation for K-shell ions at most astrophysical densities.  For the
  recombined ion, however, the level populations $N_u$ (dimensionless) are
  computed explicitly. Thus $S_{ul}$ is simply the photon line power per
  ground state of the recombining ion. For a transition with energy $E_{ul}$,
  $S_{ul}$ is related to the conventional recombination line power $P_{ul}$
  (erg $\rm{cm^3~s^{-1}}$) by
  $P_{ul}(T)=(n_H/n_e)A_Zf_{i+1}(T)S_{ul}(T)E_{ul}$, where $A_Z$ is the
  elemental abundance and $f_{i+1}$ is the ionic fraction of the recombining
  ion. The quantities $n_H$ and $n_e$ are the hydrogen and electron densities,
  respectively.

  Operationally, $S_{ul}$ is found directly from the inversion of the rate
  matrix according to $S_{ul}(T) = N_u (T) A_{ul}/n_e$, where $ A_{ul}$ is the
  radiative decay rate corresponding to the transition $u \rightarrow l$.  The
  rate matrix is solved over a grid of temperatures, and, for each X-ray
  transition in the model, $S_{ul}$ is fit to a power law; $S_{ul}(T) =C_{ul}
  ~T^{-\gamma_{ul}}$, which provides an adequate description for our
  purposes. Values of $\gamma_{ul}$ lie in the approximate range 0.6 -- 0.8.

  In constructing the spectral model, a number of physically plausible
  approximations are invoked.  First of all, we do not include line flux
  contributions from collisional excitations since we expect the local
  temperatures to be much lower than the ionization potentials ($kT/\chi \ll
  1$).  Collisional transfers in the He-like ions are also not included.  This
  approximation is valid in the low density limit ($n \lesssim 10^{10}
  ~\rm{cm^{-3}}$) and, in any case, the {\small {\it ASCA}} {\small CCD}s do
  not have the resolving power to separate these lines to observe variations
  of line ratios in He-like ions.  We also do not take into account
  inner-shell ionization of Li-like ions, which can lead to the production of
  the forbidden line in He-like ions.

  Although they are available in our spectral model, we do not include the
  forest of iron L transitions in our analysis of the Vela X-1 spectrum.  For
  X-ray photoionized plasmas, the line power of individual iron L lines is low
  compared to those of H-like and He-like ions of lower $Z$ elements which
  exist at similar ionization parameters \markcite{kall96}(Kallman et al.\
  1996).  In addition, at {\small {\it ASCA}} {\small SIS} resolution, the
  iron L lines are not resolvable -- they produce a more-or-less smooth
  continuum component between 0.7 and 2.0 keV, which provides only a minor
  perturbation to the true continuum in the same energy range.  As we show
  later, the omission of these features has a negligible effect on our derived
  wind for the Vela X-1 system.

  As indicated earlier, the line power of a particular transition produced by
  recombination cascade depends on temperature through the temperature
  dependence of the recombination coefficient.  In principle, the temperature
  appropriate to the relevant charge state can be inferred from the shape of
  the associated {\small RRC} feature, as was done using {\small {\it ASCA}}
  data for Cyg X-3 \markcite{lied96}(LP96) and 4U 1626-67
  \markcite{ange95}(Angelini et al.\ 1995).  However, for the Vela X-1 data
  set, the {\small RRC} are blended with other features, and the spectra are
  of somewhat lower statistical quality, so this is not practical in this
  case.  Instead, we must assume a temperature for each ionic species.  A
  natural choice for the temperature corresponds to that where line powers of
  the brightest lines for that ion have their maximum.  A large fraction of
  the total ionic line emission originates near this temperature.  Therefore,
  for each ion, we fix the temperature at the value where the product
  $f_{i+1}(T) S_{ul}(T)$ peaks. In photoionized plasmas, however, the charge
  state distribution of the gas is determined not by the temperature, but by
  the ionization parameter, $\xi=L_x/(n_pr_x^2)$, where $L_x$ is the X-ray
  luminosity, $n_p$ is the proton number density, and $r_x$ is the distance
  from the ionizing source \markcite{tart69}(Tarter, Tucker, \& Salpeter
  1969).  We use {\small XSTAR} to calculate fractional ionic abundances,
  $f_{i+1}$, and temperatures, $T$, as functions of $\xi$. For each ion in the
  model, we then use the $T(\xi)$ relationship, and fix the electron
  temperature according to the value of $\xi$ for which $f_{i+1}(\xi)
  S_{ul}(\xi)$ attains its maximum.  For each charge state $i$, we define the
  {\it ionization parameter of formation} ($\xi_{\rm{form}}$) to be that value
  of $\xi$ corresponding to the maximum of $f_{i+1} S_{ul}$.  We assign the
  {\it temperature of formation} according to
  $T_{\rm{form}}=T(\xi_{\rm{form}})$.

  {\small XSTAR} requires an input shape for the ionizing continuum X-ray
  spectrum.  At energies below 10 keV, this is constrained by the shape of the
  continuum visible in the {\small {\it ASCA}} spectrum itself, but the
  ionization and temperature structure is also sensitive to the shape of the
  spectrum at higher energies, outside of the {\small ASCA} bandpass.  We
  assume the spectrum determined for Vela X-1 using {\small {\it HEAO 1}} data
  by White, Swank, \& Holt \markcite{whit83}(1983), which involves a broken
  power law with an exponential cutoff at 20 keV with an e-folding energy of
  16 keV.  In Table 1, we list the $\xi_{\rm{form}}$ and their respective
  $T_{\rm{form}}$ values for the H-like and He-like species of oxygen, neon,
  magnesium, silicon, sulphur, and iron, given this assumed ionizing spectrum.

\placetable{table1}

\subsubsection{Discrete Fluorescent Emission}

  We do not explicitly model the atomic physics of fluorescent line
  production, but rather, represent them as isolated gaussian components.
  Apart from the iron K line complex in which the characteristic line width
  can be determined from the data, the mid-$Z$ fluorescent line widths are
  fixed at $\sigma = 50$ eV.  Their centroid energies and normalizations are
  left as free parameters.  The lines widths were chosen to represent typical
  spreads in fluorescent line energies for the low charge states of each
  element.  In fact, the line fluxes are not sensitive to our choice of
  $\sigma$ as long as they lie within realistic values.

\subsection{Results of Spectral Fitting}

  Spectral fits were performed to the {\small SIS}0 and {\small SIS}1 data
  during eclipse phase with {\small XSPEC}.  We have modified the code to
  incorporate the specific spectral model described in \S2.2.2.  Parameters
  for the fit include the normalizations, absorbing column densities, and
  power law indices for both the direct and scattered continuum components,
  the emission measures (proportional to the normalization of the
  recombination line spectrum and associated {\small RRC}) of the H-like and
  He-like charge states of oxygen, neon, magnesium, silicon, sulfur, argon,
  calcium, and iron and the normalizations and line centroid energies of the
  K-shell fluorescent features associated with magnesium, silicon, sulfur,
  argon, calcium, iron, and possibly nickel.  Most of these are left ``free''
  to be constrained by the fit, however, we forced the power law index of the
  scattered component to agree with that of the direct component, and applied
  the attenuation associated with the derived absorbing column density of the
  scattered component of the line and {\small RRC} radiation as well.  The
  latter is justified if both the scattered continuum and all of the discrete
  features originate in the same extended stellar wind, as we have assumed.

  Despite the large number of free parameters, there are still many (262)
  independent degrees of freedom.  Nevertheless, we find an excellent fit with
  this model with $\chi^2_r = 0.88$.  The data and best-fit model are shown in
  Figure 1, along with the residuals.  Note that all of the subtle discrete
  structure in the spectrum is accurately reproduced.

\placefigure{fig1}

\placetable{table2}

  The derived line fluxes and centroid energies for the fluorescent features
  are listed in Table 2.  The detection of many of these features was
  previously reported by Nagase et al.\ \markcite{naga97}(1997).  The centroid
  energies suggest that most of this emission arises in material that is
  partially ionized, as indicated by the likely ion identifications listed in
  the second column of the table.  Partial ionization is indeed expected for
  even dense components of the wind, given the intense ultraviolet field of
  the companion star which irradiates this material.  The reality of each
  fluorescent feature can be assessed by the change in $\chi^2$ which results
  when it is omitted from the fit, as listed in the fourth column of Table 2.
  Except for the nickel line at 7.52 keV which is required at a 91.7\%
  confidence level, all of the features are highly significant ( $>$ 99.9\%
  confidence level).

  Due to significant line blending at {\small {\it ASCA}} resolution, there is
  ambiguity in the derived fluxes of the fluorescent lines.  While the sulfur,
  argon, and calcium fluorescent features are well-separated from the bright
  recombination lines and {\small RRC} from the more highly ionized material,
  the magnesium and silicon lines are blended.  For example, the magnesium
  line at 1.30 keV lies close to Mg {\small XI} K$\alpha$ (1.34 keV) and the
  Ne {\small X RRC} edge (1.36 keV).  If we allow the temperatures of Ne
  {\small X} and Mg {\small XI} to be additional free parameters in the
  spectral fit, the necessity for the 1.30 keV feature can be removed by
  lowering the temperature of Ne {\small X}, which has the effect of
  increasing the contrast in the {\small RRC} with respect to the
  corresponding lines.  However, the required temperatures we obtain from this
  procedure (1.1 eV for Ne {\small X} and 5.3 eV for Mg {\small XI}) are
  factors of $\sim 100$ and $\sim 4$ lower than what would be expected from
  photoelectric heating in photoionization equilibrium at this level of
  ionization.  Since these temperatures are unphysical, we choose to fix the
  temperatures at their respective temperatures of formation to derive the
  fluorescent line fluxes and centroid energies.

  Similar spectral fitting procedures were performed for the {\small GIS}2
  data during the pre-eclipse and post-eclipse phases.  Our goal, in these
  cases, was merely to characterize the continuum so as to better constrain
  the ionizing spectrum and luminosity.  Thus, although we include
  contributions from discrete emission in the spectral fit, we do not attempt
  to quantify the detailed characteristics of the line spectrum, which are
  poorly determined anyway, given the lower spectral resolution of the {\small
  GIS}.  The best-fit continuum parameters for the three phases are listed in
  Table 3.  The derived power law indices agree well with those determined by
  previous experiments.  Note that the absorbing column density inferred for
  the scattered component of $(5 - 10) \times 10^{21} ~\rm{cm^{-2}}$ is
  roughly consistent with the interstellar reddening to this source observed
  in the optical and ultraviolet \markcite{nand75, cont78}(Nandy, Napier, \&
  Thompson 1975; Conti 1978).  Adopting a distance to Vela X-1 of 1.9 kpc
  \markcite{sada85}(Sadakane et al.\ 1985), we derive an average X-ray
  luminosity, which we define as the continuum luminosity above $E = 1$ Ryd,
  of $L_x = 4.5 \times 10^{36} ~\rm{erg} ~\rm{s}^{-1}$.

\placetable{table3}

\subsection{Derivation of the Empirical DEM}

  Since radiative recombination is a two-body process, the emissivity of each
  recombination line and {\small RRC} feature is proportional to density
  squared: $j_{ul}(\xi) = n_e^2 ~P_{ul}(\xi)$, where $P_{ul}$ is the line
  power, defined earlier in \S2.2.2.  If we define the emission measure to be
  $EM = \int n_e^2~dV$, the line luminosity produced at the source over a
  narrow range of $\xi$ (or $\log \xi$) --- the differential line luminosity
  --- can be expressed as $dL_{ul} = P_{ul}(\xi)~[d(EM)/d\log \xi]~d\log \xi$.
  In {\small HMXB}s, and probably most accretion powered sources, the observed
  spectrum is likely to be a superposition of individual spectra, each of
  which can be characterized by a single ionization parameter, provided that
  radiative transfer does not deform or attenuate the ionizing spectrum.
  Therefore, the line luminosity that one would infer from a measurement of
  the line flux is given by
\begin{equation}
  L_{ul} = \int d\log \xi~\biggl[
  \frac{d(EM)}{d\log \xi}\biggr]~P_{ul}(\xi).
\end{equation}
  The bracketed quantity is known as the {\it differential emission measure}
  ({\small DEM}) distribution.  For each value of $\xi$, the {\small DEM} acts
  as a weighting factor that determines the contributions of each ionization
  zone to the total line flux.  Spectroscopic analysis in this context is
  analogous to the use of a temperature-dependent {\small DEM} distribution in
  interpreting spectra from plasmas in coronal equilibrium. {\small DEM}
  analyses are used widely for modeling the structure of stellar coronae, for
  example, and have proven to be an effective means by which to extract their
  global properties \markcite{jord75}(Jordan 1975).

  If we insert the expression for $P_{ul}$ in terms of $S_{ul}$ (\S2.2.2) into
  equation (2), we have
\begin{equation}
  L_{ul} = E_{ul}~\int d\log \xi~
      \left[\frac{d(EM_{i+1})}{d\log\xi}\right]~S_{ul}(\xi),
\end{equation}
  where $d(EM_{i+1}) = n_e n_{i+1}\hspace{0.03in}dV$, and where we have used
  $n_{i+1} = (n_{H}/n_{e}) f_{i+1} A_{Z}n_{e}$. Operationally, the set
  $EM_{i+1}$ is determined from fitting to the Vela X-1 spectrum.  No
  constraints are applied to the chemical abundances or to the charge state
  distribution. These factors, as well as the electron density, are subsumed
  by $EM_{i+1}$. It is important to note, however, that each recombination
  line has a definite temperature-dependent relationship to every other line,
  as well as the {\small RRC}, emitted by that ion, and that this is fully
  accounted for in the model.

  In determining the emission measures from the spectral fit, we make a simple
  approximation that the line power of the brightest line in each ion is a
  flat function of $\log \xi$, over a given range $\Delta \log \xi$.  For most
  H-like and He-like ions, $\Delta \log \xi \sim 1$.  We plot the derived
  empirical {\small DEM} distribution as determined from the H-like and
  He-like lines and {\small RRC} of oxygen, neon, magnesium, silicon, sulfur,
  and iron.

\placefigure{fig2}

  For a given line flux, the derived {\small EM}s are relatively weak
  functions of the assumed temperature; $EM_{i+1} \propto T^{\gamma}$, where
  $\gamma \sim 0.7$.  Nevertheless, there is some arbitrariness in the
  construction of the empirical {\small DEM} distribution, and we estimate
  that our derived {\small DEM} values may be uncertain by up to a factor of 2
  or 3.

\section{Physical Model of the Ionized Stellar Wind}

  The empirical {\small DEM} distribution derived in \S2.4 is useful for
  guiding the interpretation of the spectrum in terms of realistic physical
  models of the ionized stellar wind in the system.  By constructing this
  distribution, we have effectively removed the ``spectroscopic physics'' from
  the problem, and produced a curve that can be directly compared to that
  expected for simple wind models.  In this section we explore the dependence
  of the {\small DEM} distribution on various wind parameters.  This is
  described in \S3.1 -- \S3.3.  Given the uncertainties inherent in the
  derivation of the empirical {\small DEM}, however, the final determination
  of the constraints on wind parameters can only come from direct comparison
  of fully self-consistent wind spectral models with the {\small {\it ASCA}}
  spectrum itself.  This is described in \S3.4.

\subsection{General Characteristics of DEM Distributions for the Ionized
            Stellar Wind in HMXBs}

  We begin by considering the simple case of a hard X-ray point source located
  near to a companion star with a spherically symmetric stellar wind.  We note
  that there are several mechanisms that may distort the wind from spherical
  symmetry: tidal forces of the compact object which may enhance the mass loss
  rate along the binary axis \markcite{frie82,stev88}(Friend \& Castor 1982;
  Stevens 1988), suppression of the wind velocity through photoionization
  \markcite{blon94}(Blondin 1994), non-static accretion
  \markcite{hoch87,taam88}(Ho \& Aarons 1987; Taam \& Fryxell 1988), and the
  presence of a bow shock and accretion wake \markcite{fryx88,blon90} (Fryxell
  \& Taam 1988; Blondin et al.\ 1990).  However, these effects, with the
  exception of the accretion wake, are relatively localized near the compact
  object.  Since our purpose here is to calculate the {\small DEM} during the
  eclipse phases ($-0.1 \lesssim \phi \lesssim \rm{+}0.1$), their effect on
  the {\small DEM} is small.  Therefore, we neglect all such effects and
  assume that the density distribution is spherically symmetric around the
  companion star.  We also ignore the presence of the extended stellar
  atmosphere.  Sato et al.\ \markcite{sato86}(1986) adopt an atmosphere in
  which the particle density decays exponentially with distance from the
  stellar surface.  This affects the {\small DEM} distribution only below
  $\log \xi \sim 1$, which is outside the range constrained by our spectral
  fits.

  We also assume that the radiation field of the compact object is isotropic,
  and we ignore transient effects that may be produced in the wind by X-ray
  pulsation.  The pulse profile of Vela X-1 is complex
  \markcite{mccl76}(McClintock et al.\ 1976), indicating that the emission is
  not strongly beamed.  Recent {\small {\it HST}} observations showed weak
  ($\sim 3\%$) pulsations in \ion{Si}{4} and \ion{N}{5} absorption lines,
  while the X-ray continuum varies by $\sim 50\%$ from the mean intensity
  \markcite{boro96}(Boroson et al.\ 1996).  Since the gas density in X-ray
  emission line regions is most likely to be lower than in {\small UV}
  emission/absorption regions, we expect the recombination timescales to be
  comparable to the Vela X-1 pulse period, as will be checked later.  The
  wind, therefore, responds to the time averaged flux, rather than the
  instantaneous flux of the compact object.  In addition, recombination
  timescales are also much shorter compared to the timescales for individual
  ions to traverse ionization zones.  For these reasons, we ignore the
  dynamical behavior of the ionization structure of the wind, so that it is
  meaningful to speak of local zones of ionization balance fixed in the
  system.

  In a stellar wind of massive young stars, the material accelerates radially
  from the stellar surface through momentum transfer due to absorption and
  scattering of {\small UV} continuum radiation in the resonance lines from
  low charge states of the abundant elements \markcite{lucy70}(Lucy \&
  Solomon, 1970).  We assume that the wind has an initial velocity, $v_0$,
  equal to the thermal velocity of the stellar atmosphere (typically $\sim 30
  ~\rm{km~s^{-1}}$), and accelerates until it reaches a terminal velocity,
  $v_{\infty}$.  Generalizing the results of {\small CAK}, we can write the
  velocity profile as,
\begin{equation}
  v(r) = v_{0}+v_{\infty} \left(1-\frac{R_{\ast}}{r}\right)^{\beta},
\end{equation}
  where $R_{\ast}$ is the radius of the companion, $r$ is the distance from
  the center of the companion, and $\beta$ is a parameter which determines the
  shape of the velocity profile.  For typical isolated {\small OB} stars, it
  has been shown that $\beta \sim 0.8$ \markcite{frie86,paul86}(Friend \&
  Abbott 1986; Pauldrach, Puls, \& Kudritzki 1986).

  To derive an expression for the ionization parameter for all points in the
  wind, we invoke the relation for mass conservation in a spherically
  symmetric flow ($\dot{M}=4\pi r^{2} \mu m_{p}n_{p}(r)v(r)=$ constant, where
  $\mu$ is the mean atomic weight).  The density can then be written in terms
  of the mass loss rate, the velocity of the wind, and the distance from the
  companion star.  Therefore, the ionization parameter near the binary system
  can be recast in terms of position and stellar wind parameters,
\begin{equation}
  \xi(r,r_{x}) = 4.3 \times 10^{2} \hspace{0.05in}
           \frac{(L_x)_{36} (v_\infty)_{8}} {\dot{M}_{-7}}
           \left(\frac{r}{r_{x}}\right)^{2}
           \left[\frac{v_0}{v_{\infty}}+
           \left(1-\frac{R_{\ast}}{r}\right)^{\beta}\right]
           \hspace{0.15in}\rm{erg~cm~s}^{-1},
\end{equation}
  where $(L_x)_{36}$ is the X-ray luminosity (in multiples of $10^{36}
  ~\rm{erg~s^{-1}}$), $(v_\infty)_8$ is the terminal velocity (in multiples of
  $10^{8} ~\rm{cm~s^{-1}}$), and $\dot{M}_{-7}$ is the mass loss rate of the
  companion (in multiples of $10^{-7} M_{\odot} ~\rm{yr^{-1}}$).  We note that
  $\xi(r,r_x)$ is insensitive to $v_0$ for $v_0 \ll v_\infty$.  The key
  parameters that determine the distribution of $\xi$ are the companion radius
  $R_\ast$, binary separation $a$, the X-ray luminosity $L_{x}$, the mass loss
  rate in the wind $\dot{M}$, the terminal wind velocity $v_{\infty}$, and the
  wind velocity profile parameter $\beta$.  We adopt the following values:
  companion radius, $R_\ast = 30 ~R_\odot$, binary separation, $a = 53.4
  ~R_\odot$ \markcite{vank95}(van Kerkwijk et al.\ 1995), X-ray luminosity,
  $L_x = 4.5 \times 10^{36} ~\rm{erg~s^{-1}}$ (as determined from the pre- and
  post-eclipse {\small GIS}2 spectra), and a terminal wind velocity, $v_\infty
  = 1700 ~\rm{km~s^{-1}}$ \markcite{dupr80}(Dupree et al.\ 1980).  These are
  listed in Table 4 together with the derived wind parameters, as described
  below.

\placetable{table4}

\placefigure{fig3}

  The ionization zones of a spherically symmetric wind, ionized by a point
  source of X-radiation offset from the center of the wind, have approximate
  bispherical symmetry (HM77). The symmetry is exact in the case of a
  constant-velocity wind.  An example of the contours of constant $\xi$ for a
  Vela X-1 wind model is shown in Figure 3.  For any particular wind model, we
  calculate $\xi(r,r_x)$, and construct a theoretical {\small DEM}
  distribution as follows.  We adopt a spherical coordinate system, centered
  on the companion, with a variable radial cell dimension increasing with
  radius like $(r-R_\ast)^2$, from an initial cell size of $\Delta r =
  10^{-7}~ R_\odot$.  In both the polar and azimuthal directions, the cells
  are divided into 720 bins.  The maximum radius of the volume is $20~R_\ast$.
  For each cell, we calculate the electron density from the wind model, and
  evaluate $\xi$, except for those cells lying in the shadow cone, where
  X-radiation from the compact object is blocked by the companion.  We
  separate $\xi$ into logarithmically spaced bins, with each bin having a
  width of $\Delta \log \xi = 0.025$.  The {\small DEM} contribution is
  evaluated for each cell, and added to the appropriate $\xi$-bin.
  Convergence to within a few percent at all values of $\log \xi$ is achieved
  with this procedure.  We refer to the {\small DEM} distribution calculated
  in this way as the {\it intrinsic DEM distribution}.

  For comparison to the empirical {\small DEM} distribution we need to take
  explicit account of the occultation of various parts of the wind by the
  companion star, which is a function of inclination angle and orbital phase.
  We refer to the phase-dependent, visible portion of the intrinsic {\small
  DEM} as the {\it apparent DEM distribution}.  The phase dependence is shown
  in Figure 4, where apparent {\small DEM}s at three orbital phases are
  plotted.  Note that the calculations are carried out for an assumed
  inclination angle of $74^\circ$. This permits a partial view of regions of
  high $\xi$ near the X-ray source, regions that would be entirely occulted if
  the inclination angle were sufficiently close to $90^\circ$.  Since regions
  inside the shadow cone and within the occulted region must be excluded when
  calculating the apparent {\small DEM} distribution, the total apparent EM
  (integrated apparent {\small DEM} distribution) reaches a minimum at $\phi =
  0.0$, when there is minimal (although non-vanishing) spatial overlap between
  these regions.  Conversely, the apparent EM reaches its maximum value when
  there is maximal overlap, i.e., when $\phi = 0.5$. This behavior is
  illustrated in Figure 4, along with the intermediate case, $\phi =
  0.25$. The {\small DEM} increases monotonically for all $\xi$ between phases
  0.0 and 0.5.

\placefigure{fig4}

\subsection{$\beta$ Dependence of the DEM Distribution}

\placefigure{fig5}

  The $\beta$-dependence of the {\small DEM} distribution is illustrated in
  Figure 5, where we have plotted the apparent {\small DEM} distribution for
  the eclipse phase for various values of $\beta$, assuming a mass loss rate,
  terminal velocity, and X-ray luminosity as given in Table 4.  For a constant
  velocity wind ($\beta = 0$), the {\small DEM} peaks sharply at the value of
  $\xi$ corresponding to the midplane ($r=r_x$) of the binary system --- the
  plane perpendicular to the line of centers at the midpoint.  We refer to the
  position of this peak as $\xi_{\mbox{\scriptsize mid}}$.  The peak owes its
  existence to a purely geometrical effect; the volume that covers a given
  range in $\xi$ has a global maximum at $\xi =\xi_{\mbox{\scriptsize mid}}$,
  and since the density is a smooth function of position, the {\small DEM}
  peaks at that value. The narrow width of the {\small DEM} distribution
  ($\Delta \log \xi \approx 0.6$) results in an emission line spectrum
  dominated by only a few ions.  The $\beta = 0$ case is the simplest, and
  does not account for the dynamical effects associated with the {\small UV}
  field of the companion.  It may, however, have some relevance to the case of
  a ``coasting'' wind, where the ions responsible for driving the wind have
  been ionized away by the combined {\small UV} and X-ray fields
  \markcite{macg82}(MacGregor \& Vitello 1982).

  For an accelerating wind, ($\beta > 0$), the wind velocity near the
  companion is significantly lower than the terminal velocity.  For a given
  mass loss rate, the density in this region is thus higher than that of the
  constant-velocity case. Since, locally, $d(EM) \propto v^{-2}$, the {\small
  DEM} increases for $\xi \lesssim \xi_{\mbox{\scriptsize mid}}$ (the
  companion half-space) as $\beta$ increases.  In fact, even for the $\beta
  =0.2$ case, which corresponds to extremely rapid acceleration, the {\small
  DEM} distribution has started to broaden, with a rapid increase in the
  {\small DEM} near the lower edge of the distribution. All traces of a peak
  in the distribution have vanished for the intermediate values of
  $\beta$. For $\beta$ in the range 0.4--0.7, the distribution resembles a
  trapezoid, with a width $\Delta \log \xi$ near 1.2. For more slowly
  accelerating winds $(\beta \gtrsim 0.8)$ a peak develops near $\log \xi
  =2.0$.  Comparison with the empirical {\small DEM} distribution depicted in
  Figure 2, suggests that the value of $\beta$ for the Vela X-1 wind lies in
  the range 0.4 -- 0.8.

  Regions with $\xi \gtrsim \xi_{\mbox{\scriptsize mid}}$ (the X-ray source
  half-space) are nearly unaffected by variations in $\beta$, since, by the
  time the material reaches the X-ray source half-space, it acquires a large
  fraction of its terminal velocity and, therefore, is nearly independent of
  the value of $\beta$. For higher values of $\xi$, say $\log \xi \gtrsim
  4.0$, the surfaces of constant $\xi$ approach perfect spheres concentric
  with the X-ray source. Moreover, in this idealized smooth wind
  approximation, the density does not vary drastically across a surface of
  constant $\xi$ in this region(i.e., $n_e$ is independent of $\beta$).
  Therefore, the asymptotic form of the intrinsic {\small DEM} distribution is
  given by $d(EM)/d \log \xi = 2 \pi \ln 10~ L_x ~(L_xn_p)^{1/2}~ \xi^{-3/2}$.

  Figure 5 also shows that the total {\small EM} (the integrated {\small DEM})
  increases monotonically with $\beta$. For this example, {\small EM} changes
  by a factor of 8 in comparing the $\beta = 0.0$ and $\beta = 1.0$
  distributions. However, for the most likely range of values, say $\beta =$
  0.4 -- 0.8, {\small EM} increases by less than a factor of three. We can use
  this fact to derive a simple way to estimate the mass loss rate, as
  discussed in the next section.

 \subsection{Total Emission Measure}

  The total emission measure is simply the volume integral of $n_e^2$ over the
  wind.  The electron density, $n_e$, is related to the mass density, $\rho$,
  by: $n_e = \kappa \rho/m_p$, where $\kappa$ is the mean number of electrons
  per nucleon.  In general, $\kappa$ depends on the local value of the
  ionization parameter.  However, for near cosmic abundances, free electrons
  come mainly from hydrogen and helium, so that if these two elements are
  fully ionized throughout the wind, then $\kappa$ is only very weakly
  dependent on position.  In that case, for a spherically symmetric smooth
  wind with a {\small CAK} velocity profile, we get
\begin{equation}
  EM = \frac{\dot{M}^2 \kappa^2}{4 \pi \mu^2 m_p^2 v_\infty^2 R_\ast}
       \hspace{0.05in} \int_0^1\frac{dx}
        {(v_0/v_{\infty} + x^{\beta})^2},
\end{equation}
  where $x$ is a dummy variable of integration.

  This is the total emission measure in the wind.  To compare with the
  integral of the empirical {\small DEM} depicted in Figure 2, we need to
  convert to an apparent emission measure $EM_{\rm{app}}$, which is related to
  $EM$ by a geometric scale factor, $f<1$, such that $EM_{\rm{app}} =f \cdot
  EM$.  If we denote the integral by $I$, equation (6) can be inverted for
  $\dot{M}$ to give
\begin{equation}
\dot{M}=2.7 \times 10^{-7}~(fI)^{-1/2}~
        \biggl(\frac{v_{\infty}}{1700 ~\rm{km~s^{-1}}} \biggr)~
        \biggl(\frac{EM_{\rm{app}}}{10^{56}~\rm{cm^{-3}}}\biggr)^{1/2}~
        M_\odot ~\rm{yr^{-1}}.
\end{equation}
  A lengthy calculation is required to determine $f$ in a general case.  For
  the Vela X-1 system during eclipse, $f \approx 0.30$.  The integral $I$
  depends on $\beta$ and the ratio $v_0/v_{\infty}$.  We find that, for
  typical values of these parameters, $I$ falls into the range $\sim$10 -- 20.
  By inspection of Fig.\ 2, a reasonable estimate for $EM_{\rm{app}}$ is $3
  \times 10^{56}~\rm{cm^{-3}}$.  Therefore, the implied mass loss rate is
  $\sim (1-3) \times 10^{-7}~M_\odot ~\rm{yr^{-1}}$.

\subsection{Self-Consistent Spectral Fit to the Physical Wind Model}

  Having determined approximate values of $\beta$ and $\dot{M}$, we can now
  work backward, and use our physical model of the wind to generate an
  explicit spectral model which can be compared with the {\small {\it ASCA}}
  eclipse phase data.  We assume a specific set of chemical abundances, a
  detailed model for the charge state distribution, and a global model of the
  wind to determine the density and temperature distribution.  For the
  chemical abundances, we take the solar photospheric set from Anders \&
  Grevesse \markcite{ande89}(1989).  The ion fractions and temperature are
  calculated for each value of $\xi$ with {\small XSTAR}, as described in
  \S2.2.2.  The density at each point in the model wind follows from
  specifications of $\dot{M}$ and $\beta$.  The model {\small DEM}
  distributions are calculated by averaging the apparent {\small DEM}s over
  the orbital phases $-0.1 \lesssim \phi \lesssim 0.1$, which corresponds to
  the eclipse phases of Vela X-1, and mimics the time integration of spectra
  accumulated during an interval of changing orbital configuration.

  The 0.0 -- 5.0 range in $\log \xi$ is divided into 200 bins of width $\Delta
  \log \xi = 0.025$, as before. At each $\xi$ on the grid, $T$ and the set
  $f_{i+1}$ are found from {\small XSTAR}.  The line powers of each
  recombination line in the model are evaluated according to our
  temperature-parameterized line power model.  The process is repeated for the
  next value of $\xi$ on the grid, and so forth, until a set of 200 spectral
  $\xi$-components have been calculated.  Each {\small DEM} model assigns a
  different weighting to each $\xi$-component; the product of this weighting
  factor and the line power constitutes the integrand of equation (3).  A sum
  over the $\log \xi$-grid gives us $L_{ul}$, as in equation (3).  Continuum
  and fluorescent lines are added to the recombination spectrum.  The model
  spectrum is modified by geometrical dilution, and by attenuation by a column
  density of neutral material, then convolved with the {\small SIS} instrument
  response, and compared to the data.  Values of $\chi^2$ are determined on a
  $50 \times 70$ grid in $\log \dot{M}$ -- $\beta$ parameter space.  The
  domains of $\log \dot{M}$ $(M_\odot ~\rm{yr^{-1}}$) and $\beta$ are $(-6.85,
  -6.34)$ and $(0.45, 1.14)$, respectively, with a grid spacing of 0.01 in
  both dimensions.

  We first perform these comparisons by fixing the continuum fluxes and column
  densities to their best-fit values of the original fit with the
  recombination cascade model (\S2.4).  With $\dot{M}$ and $\beta$ left as
  free parameters, we find that $\chi^2$ attains its minimum for $\dot{M} =
  2.5^{+0.6}_{-0.5} \times 10^{-7} ~M_\odot ~\rm{yr^{-1}}$ and $\beta =
  0.65^{+0.16}_{-0.18}$, where the errors correspond to 90\% confidence ranges
  for two interesting parameters.  Although the fit is statistically
  acceptable ($\chi^2_r = 1.27$ for 297 degrees of freedom), a better fit is
  obtained if the column density of the scattered component is allowed to
  vary.  The data are then fit with three free parameters: $\dot{M}$, $\beta$,
  and $N_H^{\rm{scat}}$. We find that the best-fit values in this case are
  $\dot{M} = 2.65^{+0.65}_{-0.50} \times 10^{-7} ~M_\odot ~\rm{yr^{-1}}$,
  $\beta = 0.79^{+0.23}_{-0.23}$, and $N_H^{\rm{scat}} = 6.8^{+0.8}_{-0.7}
  \times 10^{21}$ cm$^{-2}$, with $\chi^2_r = 1.01$ for 296 degrees of
  freedom. The errors here correspond to 90\% confidence ranges for three
  interesting parameters.  The 90\% and 99\% confidence contours for the two
  wind parameters, $\dot{M}$ and $\beta$, are plotted in Figure 6, together
  with some previous determinations using other techniques.

\placefigure{fig6}

  The best-fit $\dot{M}$ found here agrees well with the estimate based upon
  the much simpler approach using equation (7). The close agreement is
  somewhat fortuitous, since our earlier estimate does not take account of the
  details of the {\small DEM} distribution. This validates our earlier claim
  that $\dot{M}$ can be determined to within factors of a few from an estimate
  of the total emission measure. Our earlier estimate for $EM_{\rm{app}}$ was
  $3 \times 10^{56}~\rm{cm^{-3}}$. From a detailed spectral analysis, our
  best-fit value is $2.63 \times 10^{56}$ $\rm{cm^{-3}}$.

  The best-fit column density is higher than the Galactic value ($4 \times
  10^{21} ~\rm{cm^{-2}}$) by roughly a factor of two.  The excess absorption
  could be ascribed to material local to the Vela X-1 system. Note that
  $\beta$ and $N_H^{\rm{scat}}$ are highly correlated; since increasing
  $\beta$ increases the {\small DEM} of low-$\xi$ material (see Figure 5), a
  higher column density can offset this effect by absorbing the low energy
  line fluxes.

  As illustrated in Figure 6, our derived range in $\beta$ is consistent with
  that inferred from {\small UV} observations \markcite{dupr80}(Dupree et al.\
  1980) and close to the value calculated by Friend \& Abbott
  \markcite{frie86}(1986), while our derived mass loss rate is at least a
  factor of four lower than previously published results, and approximately an
  order of magnitude lower than the midrange of typically quoted values (1 --
  7) $\times ~ 10^{-6}~M_\odot ~\rm{yr^{-1}}$
  \markcite{hutc76,dupr80,kall82,sada85,sato86}(Hutchings 1976; Dupree et al.\
  1980; Kallman \& White 1982; Sadakane et al.\ 1985; Sato et al.\ 1986).  The
  best-fit model parameters are summarized in Table 4, and the best-fit
  {\small DEM} curve is shown in Figure 7.  Also, for completeness, the
  best-fit spectrum and residuals using the three-parameter model are shown in
  Figure 8.

\placefigure{fig7}

\placefigure{fig8}

  It should be noted, however, that our derived mass loss rate depends on the
  assumed metal abundances.  If we assume that the abundances are 1/10 that of
  solar, the inferred mass loss rate will be higher by a factor of $\sqrt{10}$
  due to the $\dot{M}^2$-dependence on the magnitude of the {\small DEM}
  curve.  Therefore, $\dot{M} = 2.7 \times 10^{-7} A_{Z_\odot}^{-1/2}$, where
  $A_{Z_\odot}$ is the abundances relative to solar.  Similarly, $\beta$ can
  be decreased if the abundances of low-$Z$ elements (O, Ne, and Mg) are
  increased.  The exact dependence of the inferred $\beta$ on $A_{Z_\odot}$ is
  more complex, because the $\beta$ dependence on the {\small DEM} curve is
  non-uniform.  For example, the {\small DEM} at $\log \xi \approx 2$ and
  $\beta = 0.8$ is a factor of $\sim 3$ higher than that of $\beta = 0.6$.
  Since an increase in {\small DEM} can be offset by an increase in the
  abundance of elements for which $\xi_{\rm{form}}$ lies near that value of
  $\log \xi$, a large deviation in the assumed low-$Z$ metallicity only
  corresponds to a small change in the derived $\beta$.

  As a cross-check on the wind parameters, we can calculate the implied
  Thomson scattering depth through the wind.  Since Thomson scattering is
  responsible for all of the scattered continuum observed during eclipse, the
  ratio of the normalizations of the scattered and absorbed continuum
  components $A^{\rm{scat}}/A^{\rm{dir}}$ gives a rough estimate of the
  average Thomson optical depth through the surrounding wind.  The ratio of
  the scattered continuum flux to that of the direct continuum is $ \sim 1.2
  \times 10^{-3}$ during pre-eclipse and $\sim 1.6 \times 10^{-3}$ during
  post-eclipse.  This is roughly consistent with the angle-averaged electron
  scattering optical depth ($\langle\tau_e\rangle_\Omega = 2.3 \times
  10^{-3}$) calculated using stellar wind parameters inferred from the {\small
  DEM} analysis.  Were $\dot{M}$ an order of magnitude larger than our
  best-fit value, the normalization $A^{\rm{scat}}$ would also have been an
  order of magnitude larger.

  In order to justify the elimination of iron L-shell ions from our spectral
  model, we calculated a model iron L-shell spectrum in accordance with the
  best-fit {\small DEM} distribution.  Assuming that the iron abundance is
  solar, we find that iron L line and {\small RRC} flux is 12$\%$ of the total
  line flux, where the total includes emission from all other elements used in
  the fit.  When folded through the {\small SIS} response, the iron L-shell
  spectrum resembles a faint continuum, owing to the high line density in
  L-shell spectra.  This introduces a small error into the normalizations of
  the true continuum flux components, but does not have a significant bearing
  on our results concerning the {\small DEM} analysis. The exclusion of iron
  L-shell ions leads to a great simplification in the spectral fit.

  Using the best-fit parameters listed in Table 4, we estimate typical
  recombination timescales of highly ionized ions in the wind.  The electron
  densities in the visible portion of the wind during eclipse are on the order
  of $n_e \sim 10^{8-9} ~\rm{cm^{-3}}$.  For a typical recombination
  coefficient of $\alpha_{\rm{RR}} \sim 10^{-11} ~\rm{cm^3~s^{-1}}$, the
  recombination timescales are longer than $\sim 100 ~\rm{s}$ which is
  comparable to the pulse period of Vela X-1 (283 s).  Since the ionizing
  emission is not strongly beamed, this justifies our approximation of an
  isotropic radiation field from the compact X-ray source.

\section{K-Shell Fluorescent Lines}

  The physical model for the stellar wind that we have used to derive a mass
  loss rate and velocity profile only accounts for highly ionized material
  (H-like and He-like ions).  As indicated, however, our spectral fits also
  require the presence of fluorescent lines from much lower charge states of
  many of the same elements.  The fluorescent lines have been incorporated
  into the model on an ad hoc basis, as required by the data.  In this
  section, we use the observed fluxes in these lines to infer the conditions
  in the reprocessing gas which is responsible for their emission.

  Fe K$\alpha$ fluorescence has been previously detected in the Vela X-1 X-ray
  spectrum with a number of earlier observatories
  \markcite{beck78,ohas84,lewi92}(Becker et al.\ 1978; Ohashi et al.\ 1984;
  Lewis et al.\ 1992), and several candidate sites for its origin have been
  posited.  Ohashi et al.\ \markcite{ohas84}(1984) noted that the iron line
  intensity during eclipse is roughly an order of magnitude lower than that at
  other binary phases, and concluded that a large fraction of this emission
  must be produced within $\sim 5 \times 10^{11} ~\rm{cm}$ of the neutron
  star.  However, in the simple undisturbed model of the wind, the local
  ionization parameter within this region is much too high ($\log\xi \gtrsim
  3$) to support near-neutral ions.  Hence, fluorescent lines must be produced
  in anomalous, dense structures in this region.  As shown by Blondin et al.\
  \markcite{blon90}(1990), disruption of the wind due to the presence of the
  X-ray source produces dense filaments in the accretion wake that may be 100
  times denser than the undisturbed wind.  Therefore, these filaments have
  local ionization parameters that are up to two orders of magnitude lower
  than the ambient wind, and are capable of supporting ions of low charge
  state.  The atmosphere of the companion is also a possible site for
  fluorescent line emission.  The atmosphere has densities as high as a few
  $\times~ 10^{12} ~\rm{cm^{-3}}$ out to a few tenths of a stellar radius from
  the photosphere \markcite{frie82}(Friend \& Castor 1982).  These regions
  have relatively low ionization parameters ($\Delta \log \xi \lesssim 1$) and
  enough column density ($N_{\rm{H}} \gtrsim 10^{23} ~\rm{cm^{-2}}$) to
  produce the observed line fluxes.  During eclipse, however, regions near the
  neutron star, as well as the irradiated face of the companion are mostly out
  of our line of sight. Thus a non-negligible fraction of the observed line
  flux must originate from more extended regions.

  Assuming that the fluorescing medium is optically thin to the ionizing
  continuum, the luminosity of a single K-shell fluorescent line with line
  energy $E_{K}$ and fluorescence yield $Y_{K}$ can be approximated by
\begin{equation}
  L_{K} = L_x\hspace{0.04in}Y_{K}\hspace{0.02in}
          G\hspace{0.04in}
          \frac{2-\Gamma}{2+\Gamma}
          \left(\frac{E_{c}}{\chi}\right)^{\Gamma-2}
          \frac{E_{K}}{\chi} ~\langle\tau(\chi)\rangle_{\Omega},
\end{equation}
  where we use the same cutoff power law as in the {\small XSTAR} calculations
  for the shape of the ionizing continuum, and a photoionization cross section
  of the form $\sigma(E)=\sigma(\chi)(\chi/E)^{3}$, where $\chi$ is the
  ionization potential.  The continuum source has an X-ray luminosity $L_x$.
  The factor $G$ is a rather complicated function that depends on $\chi$,
  $E_c$, $E_f$, and $\Gamma$. For a given spectrum, it depends only on $\chi$.
  The final factor in equation (8) is the solid-angle averaged photoionization
  optical depth at threshold. The optical depth here refers to paths from the
  X-ray source to points in the surrounding wind.  Explicitly,
  $\langle\tau(\chi)\rangle_{\Omega}$ is given by
\begin{equation}
  \langle\tau(\chi)\rangle_{\Omega} = \sigma(\chi)~
        \int_{\Omega}\frac{d\Omega}{4\pi}\hspace{0.05in}N_{Z},
\end{equation}
  where $N_Z$ is the column density of a given element, i.e., a sum over all
  charge states of a given element. The evaluation of equation (9) is
  simplified by assuming that $N_Z$ is angle-independent. The integral then
  becomes $N_Z \Delta \Omega/4\pi$, where $\Delta \Omega/4\pi$ is the covering
  fraction of the fluorescing material with respect to the X-ray source.

  We use fluorescent yields from Kaastra \& Mewe (1993)\markcite{kaas93}, and
  photoionization cross sections from Verner et al.  (1996)\markcite{vern96}
  to calculate the values of $N_{Z}$ required to produce the observed
  fluorescent line fluxes.  We define $N_{\rm{H}}^{\mbox{\scriptsize equiv}}$
  to be the hydrogen column density of the region which produces the
  observable fluorescent lines (the ``equivalent'' hydrogen column density),
  such that $N_Z = A_Z~N_{\rm{H}}^{\mbox{\scriptsize equiv}}$. In evaluating
  these quantities, we assume solar abundances.  For the ionizing spectrum of
  Vela X-1, we find that $G=0.937$, which is accurate to within $1\%$ for all
  elements used in our spectral model.

\placefigure{fig9}

  We have used equations (8) and (9) and the measured fluorescent line fluxes
  to infer values of $N_{\rm{H}}^{\mbox{\scriptsize equiv}} (\Delta
  \Omega/4\pi)$.  The results are plotted in Figure 9 versus the atomic number
  $Z$ of the element involved.  As can be seen, there is a clear, monotonic
  dependence on $Z$.  In particular, the derived value of
  $N_{\rm{H}}^{\mbox{\scriptsize equiv}} (\Delta \Omega/4\pi)$ is a factor of
  $\sim 20$ higher for iron ($Z = 26$) than it is for magnesium ($Z = 12$).
  This obvious trend may be a result of a process known as resonant Auger
  destruction \markcite{ross96}(Ross, Fabian, \& Brandt 1996), which can be
  significant for low-$Z$, L-shell ions in regions of moderate line optical
  depth.  If the optical depth is significant, a K-shell fluorescent line
  emitted by one atom can be resonantly absorbed by another as long as there
  is a vacant hole in the L-shell.  Since the autoionization yield for low-$Z$
  ions is high in comparison to the fluorescence yield, the second excited
  atom will preferentially Auger decay, thereby destroying the original
  fluorescence photon.  Even line optical depths of only a few can almost
  entirely suppress fluorescent K$\alpha$ emission for such species.  The
  required column densities are typically $\sim 10^{20} ~\rm{cm^{-2}}$, which
  is well below that inferred from our fluorescent line measurements.

  Resonant Auger destruction is not effective for M-shell ions, where the
  L-shell is entirely filled and resonant K-shell line absorption is
  prohibited.  However, the presence of such very low charge states may be
  suppressed by the intense {\small UV} field of the companion star, {\small
  HD} 77581, in the Vela X-1 system ($T_{\rm{eff}} = 26,000 ~K$; $L = 5 \times
  10^5 ~L_\odot$; \markcite{hutc74}Hutchings 1974), which can effectively
  ionize all atoms in the wind with ionization potentials less than $\sim 50$
  eV \markcite{kall82a}(Kallman \& McCray 1982).  The low-$Z$ elements: neon,
  magnesium, silicon, and sulfur should exist almost entirely in L-shell
  species due to this effect, where resonant Auger destruction is effective.
  By contrast, the {\small UV} field is not sufficiently intense to fully
  strip the M-shell of higher-$Z$ elements such as calcium and iron.  These
  combined effects at least qualitatively account for the lower values of
  $N_{\rm{H}}^{\mbox{\scriptsize equiv}} (\Delta \Omega/4\pi)$ inferred for
  magnesium, silicon, and sulfur, and the complete absence of fluorescent
  emission from neon.  If this interpretation is correct, then the true column
  density of cold reprocessing gas in the wind is probably closest to that
  inferred for iron, $\gtrsim 10^{22} ~\rm{cm^{-2}}$.

  As a final comment, note that there are two values plotted for iron in
  Figure 9, one derived from the Fe K$\alpha$ line, and the other from the Fe
  K$\beta$ line.  The two incommensurate values of
  $N_{\rm{H}}^{\mbox{\scriptsize equiv}} (\Delta \Omega/4\pi)$ arise from the
  apparently anomalous K$\beta$/K$\alpha$ ratio (0.28$^{+0.10}_{-0.08}$)
  compared to the expected value of 0.13 \markcite{kaas93}(Kaastra \& Mewe
  1993) under most conditions.  A similar anomaly was reported by White,
  Kallman, \& Angelini \markcite{whit97}(1997) for the {\small {\it ASCA}}
  spectrum of the low-mass X-ray binary 4U 1822 -- 37.  It should be noted,
  however, that the fluorescent emission line regions are optically thin to
  both Thomson scattering and Fe K-edge absorption as evident from the
  inferred equivalent hydrogen column density.  The spectrum of the primary
  continuum radiation reflected from this medium does not produce an
  appreciable Fe K-edge structure and, hence, does not contaminate the Fe
  K$\beta$ fluorescent line.  Therefore, we believe that this anomalous
  K$\beta$/K$\alpha$ ratio is real and the origin of this effect is still
  mysterious.

\section{Discussion}

  We have shown that the {\small {\it ASCA}} eclipse phase spectrum of Vela
  X-1 is quantitatively well described by a simple spectral model
  incorporating direct and scattered continuum radiation from the compact
  X-ray source, hydrogenic and helium-like recombination lines and their
  associated {\small RRC} from a range of intermediate-$Z$ elements (oxygen
  through iron), and fluorescent line emission from near-neutral species of
  many of the same elements (magnesium through iron).  The derived ion
  emission measures for the ionized component were assembled into an empirical
  {\small DEM} distribution, which was shown to closely resemble predicted
  {\small DEM} distributions from a simple, spherically symmetric {\small CAK}
  model of a stellar wind, ionized by a compact X-ray source offset from the
  center of the wind.  We then showed that this same physical model of the
  wind can be used to generate a self-consistent spectral model which also
  provides a remarkably good fit to the {\small {\it ASCA}} eclipse-phase
  spectrum.  The derived mass-loss rate for the wind from this analysis is
  $\dot{M} = 2.65^{+0.65}_{-0.50} \times 10^{-7} ~M_\odot ~\rm{yr^{-1}}$, and
  the derived velocity profile parameter is $\beta = 0.79^{+0.23}_{-0.23}$.

  Although our derived velocity profile is in good agreement with previous
  estimates, our derived mass loss rate is roughly a factor 10 lower than
  typical values quoted in the literature for this source.  For example,
  Dupree et al.\ \markcite{dupr80}(1980) and Sadakane et al.\
  \markcite{sada85}(1985) found $\dot{M} \sim 1 \times 10^{-6} ~M_\odot
  ~\rm{yr^{-1}}$ by measuring P~Cygni line profiles of Si {\small IV} and C
  {\small IV} resonance lines during eclipse, and comparing them to
  theoretical predictions.  However, it should be noted that the fractional
  abundance of both Si {\small IV} and C {\small IV} is quite low throughout
  the wind in their models.  Thus their inferred mass loss rates are strong
  functions of their assumed ionization fractions on the ``tails'' of the
  charge state distribution curves, which makes them quite sensitive to the
  assumptions of the model.

  Mass loss estimates have also come from X-ray absorption measurements close
  to eclipse egress.  For example, Kallman \& White \markcite{kall82}(1982)
  found $\dot{M} = (2 - 4) \times 10^{-6} ~M_\odot ~\rm{yr^{-1}}$, based on
  column densities inferred from {\it Einstein} Solid State Spectrometer
  spectra in the phase interval $\phi = 0.19 - 0.22$. However, the observed
  phase dependence of the absorbing column density is incompatible with the
  profile expected from simple stellar wind models.  In general, the observed
  column densities are high ($N_H \sim 10^{24} ~\rm{cm^{-2}}$) right after
  eclipse egress ($\phi \sim 0.1$), then decrease rapidly until $\phi \sim
  0.2$, then increase slowly throughout the remainder of the orbit.  In
  contrast, smooth stellar wind models predict a steep decrease in absorbing
  column density only within a narrow phase range, $0.10 < \phi < 0.12$.  Sato
  et al.\ \markcite{sato86}(1986) invoked an extended stellar atmosphere for
  the companion star to explain this behavior.  In their model, the wind
  contribution to the absorption is small compared to that of the atmosphere
  for phases $0.10 < \phi < 0.15$.  However, they still require a mass loss
  rate $\sim 10^{-6} ~M_\odot ~\rm{yr^{-1}}$ in order to get their minimum
  observed column density, which is $\sim 10^{22} ~\rm{cm^{-2}}$,
  significantly above that expected for the interstellar contribution alone.

  Our own best-fit wind model for the {\small {\it ASCA}} eclipse phase data
  also requires an absorbing column density of $\sim 7 \times 10^{21}
  ~\rm{cm}^{-2}$, which is higher than the canonical interstellar value.
  Since the Vela X-1 system is highly inclined ($i > 74^{\circ}$; Conti
  \markcite{cont78}1978), it is possible that the excess absorption is
  specific to the equatorial plane, where the stellar wind is preferentially
  enhanced \markcite{blon90}(Blondin et al.\ 1990).  However, it seems
  unlikely that a factor of 10 discrepancy in the mass loss rate can be due to
  that effect alone, especially during eclipse phase, where emission from the
  vicinity of the equatorial plane is largely suppressed by occultation.

  It is interesting to compare Figure 3 with Figure 9 of Dupree et al.\ (1980)
  and Figure 1 of McCray et al.\ (1984), which show values of $\log \xi$ near
  the Vela X-1 midplane close to 1.0 rather than 3.0.  The higher mass loss
  rates deduced from the {\small UV} observations translate directly into
  larger emission measure at low ionization parameters, and are {\it
  fundamentally incompatible} with the observed {\small {\it ASCA}} eclipse
  phase spectra for smooth stellar wind models.  Our calculations show that
  the {\small DEM} distribution at $1.8 < \log \xi < 2.0$, for example, would
  exhibit emission measures that are approximately two orders of magnitude
  larger than derived, which means the expected intensities of \ion{O}{8},
  \ion{Mg}{11}, and \ion{Si}{13} recombination lines would be 100 times larger
  than actually observed.

  This conflict between the previously derived mass loss rates and our own
  estimate based on the highly ionized component of the X-ray spectrum is
  exacerbated by the presence of fluorescent emission in the {\small {\it
  ASCA}} data.  Fluorescent emission is the dominant line emission process in
  lower charge states.  However, inspection of Figure 3 indicates that in our
  smooth wind model there is insufficient low-$\xi$ ($\log \xi \lesssim 1$)
  material in the observable portions of the Vela X-1 system to produce these
  lines.  Therefore, irrespective of the details of the fluorescent lines, it
  is clear that their mere existence in the eclipse spectrum signals a problem
  with the simple wind model used to describe the recombination spectrum.

  One way to reconcile the coexistence of highly ionized and near-neutral ions
  is to consider inhomogeneous stellar wind models, where cool, dense clumps
  of material are embedded in a more tenuous, highly ionized wind.  Such
  inhomogeneities are, in fact, expected theoretically for line-driven stellar
  winds, as a result of various instabilities
  \markcite{macg79,lucy80,carl80,abot80,owoc84}(MacGregor, Hartmann, \&
  Raymond 1979; Lucy \& White 1980; Carlberg 1980; Abbott 1980; Owocki \&
  Rybicki 1984).  For example, Rayleigh-Taylor instabilities should produce
  clumps with sizes $\sim 10^{11}$ cm at intermediate distances from the star
  \markcite{carl80}(Carlberg 1980).  There is also observational evidence for
  the existence of clumps in the wind.  Nagase et al.\ \markcite{naga86}(1986)
  suggested that partial obscuration, produced by clumps with typical column
  densities $\sim 10^{22} ~\rm{cm^{-2}}$, could produce the soft excess
  observed at particular orbital phases in Vela X-1.  White, Kallman \& Swank
  \markcite{whit83b}(1983) invoked overdense clumps with characteristic sizes
  $\sim 10^{11} ~\rm{cm}$ to account for observed short-term variability in
  luminosity and absorbing column density for the similar {\small HMXB} system
  4U 1700 -- 37.  In that context, it is interesting to note that the second
  {\small {\it ASCA}} observation of Vela X-1 performed in 1995 exhibited a
  flare lasting $\sim 10^4 ~\rm{s}$, in which the X-ray luminosity increased
  by as a much as a factor of three.  It is conceivable that this flare was
  the result of an increase in the accretion rate resulting from a collision
  of the neutron star with such an overdense region of the wind.  Since the
  orbital velocity of the neutron star in Vela X-1 is $\sim 300 ~\rm{km}
  ~\rm{s^{-1}}$, the required size of this region is again of order $10^{11}
  ~\rm{cm}$.

  The presence of cool, low ionization clumps in the wind can be
  quantitatively constrained by our observed fluorescent line intensities.  If
  mass is being lost by the companion star in the form of low ionization
  clumps at a rate $\dot{M}_c$, then for a spherically symmetric outflow of
  these clumps at a constant velocity, $v$, the solid angle subtended by
  clumps as seen from the compact X-ray source is approximately given by
  $\Omega = \pi \dot{M}_c r_c^2/(m_c v R_\ast$), where $r_c$ is the typical
  radius of a clump and $m_c$ is a typical mass.  For a uniform density, the
  mass of a clump is given by $m_c = (4 \pi /3) r_c^3 \mu m_p n_p$, where
  $n_p$ is the proton density within the clump, and $\mu$ is the mean atomic
  weight.  These last two expressions can be combined to give
\begin{equation}
\dot{M}_c \approx 2 \times 10^{-6}~
          \biggl(\frac{N_c\Omega/4\pi}{10^{22}~\rm{cm^{-2}}}\biggr)~
          \biggr(\frac{v_{\infty}}{1700~\rm{km~s^{-1}}}\biggr)~
          M_\odot ~\rm{yr^{-1}},
\end{equation}
  where $N_c = n_c r_c$ is the column density of a single clump, and we have
  taken the value for $R_\ast$ in Table 4.  Equating $N_c\Omega/4\pi$ to the
  values of $N_{\rm{H}}^{\rm{equiv}}$ plotted in Figure 9 for the high-$Z$
  fluorescent lines (where resonant Auger destruction is ineffective; see \S
  4), we infer mass loss rates in this clumped component which are now quite
  commensurate with previous estimates.

  The fact that the fluorescent lines are produced in near-neutral species
  imposes an upper limit, $\xi_{\rm{max}}$, on the ionization parameter in the
  clumps, which, for a given distance, implies a lower limit on the clump
  particle density: $n_p > 3 \times 10^{11} \xi^{-1} (r_x/a)^{-2}
  ~\rm{cm^{-3}}$, where $r_x$ is the distance from the X-ray source and $a$ is
  the binary separation.  For a clump column density of $\sim 10^{22}
  ~\rm{cm^{-2}}$, the implied value of $r_c$ is $\sim 10^{11} ~\rm{cm}$, in
  good agreement with theoretical expectations.  In addition, at the required
  $\xi_{\rm{max}}$, the gas temperature within the clump is $\sim$ few eV,
  which means that for the inferred values of $n_p$, the clumps are in near
  pressure equilibrium with the lower-density, higher-temperature wind.

  As demonstrated, the clumped component of the wind accounts naturally for
  the observed fluorescent lines and for the variability in the luminosity and
  the inferred absorbing column density.  Since it enhances the column density
  in low ionization material, it also helps to explain the strength of the low
  charge state absorption lines (C {\small IV} and Si {\small IV}) measured in
  the {\small UV}.  A concern might be raised about the apparent smoothness of
  the observed P~Cygni profiles in these lines: since individual clumps should
  have well-defined velocities, the lines might appear ``choppy'' if only a
  few clumps contribute to each absorption line at any particular time
  (sometimes called ``discrete absorption components'').  However, a simple
  estimate shows that the number of clouds per velocity interval is
  approximately given by:
\begin{equation}
\frac{dC}{dv}  \approx \frac{\dot{M}_cR_{\ast}}{m_c v_{\infty}^2}
               =0.7 ~(\dot{M}_c)_{-6}~(n_c)_{11}^{-1}~
               (r_c)_{11}^{-3}~
               \biggl(\frac{v_{\infty}}{1700~\rm{km~s^{-1}}}\biggr)^{-2}
               ~~\rm{(km~s^{-1})^{-1}},
\end{equation}
  which leads to approximately one clump per 1 -- 2 km s$^{-1}$ velocity
  interval.  The {\small {\it IUE}} high-dispersion spectra of Vela X-1 had a
  velocity resolution $\sim 20 ~\rm{km~s}^{-1}$, which suggests that the
  observed profile should indeed be smooth.  The fraction of the companion
  disk which should be covered by the clump ensemble is approximately:
\begin{equation}
f_{disk}=g \hspace{0.05in} \frac{\dot{M}_c}{m_p v_{\infty} N_c R_{\ast}}
        =0.6~(\dot{M}_c)_{-6}~(N_c)_{22}^{-1}~
        \biggl(\frac{v_{\infty}}{1700~\rm{km~s^{-1}}}\biggr)^{-1}.
\end{equation}
  where $g$ is a numerical factor of order 1/10.  Thus, the large covering
  factor with respect to the companion appears to be easily attainable in this
  picture.  The presence of the clumped component is also supported by the
  orbital phase variability of the P~Cygni lines which show evidence of
  non-monotonicity in the velocity structure of the wind
  \markcite{kape93}(Kaper, Hammerschlag-Hensberge, \& van Loon 1993).

  Given our estimates derived above, the clumped component of the wind
  dominates the mass loss rate from the companion.  Nevertheless, the hot
  ionized component dominates the volume of the wind.  Within an outer radius
  comparable to the binary separation, we find that the volume filling factor
  of the clumps is only $\sim 0.04$.  Therefore, the presence of the clumped
  component only mildly perturbs the otherwise smooth wind of ionized gas, and
  does not significantly affect our {\small DEM} analysis.

  This interpretation of the Vela X-1 spectrum can be quantitatively tested
  with much higher precision when high resolution X-ray spectra become
  available from {\small {\it AXAF, XMM}}, and {\it Astro-E}.  At the
  resolution expected for the spectrometers on these missions, the iron L
  spectrum will be especially constraining.  The iron L ions sample a large
  range in $\xi$ ($\log \xi = 1.8 - 2.8$) which roughly brackets the {\small
  DEM} distribution inferred from our {\small {\it ASCA}} observations.  The
  advantage of using iron L lines is that only a single elemental abundance
  must be assumed in establishing the scale for the {\small DEM} distribution.
  In the case of the {\small {\it ASCA}} data, where only hydrogenic and
  helium-like lines are available, the actual shape of the inferred {\small
  DEM} distribution is affected by assumed abundances.

  Future observations will allow the K$\alpha$ fluorescent line complexes to
  be resolved into their individual charge state contributions.  This may
  allow us to study the acceleration zone of the wind, the atmosphere of the
  companion, and the formation and evolution of the dense structures that we
  have postulated here.

  We have shown that the orbital phase variations of the apparent {\small DEM}
  can be dramatic for some ranges of $\xi$.  Observations made at different
  orbital phases will provide the information needed to refine the simple wind
  model and to describe the extended components of the wind.  Substructures
  such as the accretion wake and the magnetosphere may reveal themselves as
  perturbations to the {\small DEM} distribution predicted by extrapolating
  our model distribution to orbital phases away from eclipse.  Line flux and
  temperature measurements extracted from orbital phase-dependent,
  high-resolution spectra, when coupled with velocity information obtained
  from Doppler shifts and broadening, will provide access to a formidable set
  of diagnostic tools.

\acknowledgements

  This work has benefited from useful discussions with David Cohen, Chris
  Mauche, and Andy Rasmussen.  We have made use of data obtained through the
  High Energy Astrophysics Science Archive Research Center Online Service,
  provided by the {\small NASA}/Goddard Space Flight Center.  We thank the
  {\small {\it ASCA}} team and the {\small HEASARC} Online Service for their
  support.  This work was supported under a {\small NASA} Long Term Space
  Astrophysics Program grant ({\small NAG}5-3541).  D. A. L. was supported in
  part by a {\small NASA} Long Term Space Astrophysics Program grant ({\small
  S}-92654-{\small F}).  Work at {\small LLNL} was performed under the
  auspices of the U. S. Department of Energy, Contract No. {\small
  W}-7405-Eng-48.

\clearpage

\begin{figure}
  \caption{ The {\small {\it ASCA}} {\small SIS}0 data and the best-fit
     recombination cascade model.  The strongest recombination and fluorescent
     lines are labeled.  {\small SIS}1 data was also used in the analysis but
     are not shown for clarity.}  \label{fig1}
\end{figure}

\begin{figure}
  \caption{ The derived emission measure distribution based on the best fit to
     the {\small {\it ASCA}} {\small SIS} data.  Emission measures from Ar and
     Ca are not shown in the figure.  The error bars correspond to 90\%
     confidence ranges for one interesting parameter.}  \label{fig2}
\end{figure}

\begin{figure}
  \caption{ Ionization contours for the parameters shown in Table 4.  The
     contours represent surfaces of constant ionization parameter.
     Axisymmetric surfaces of revolution can be derived by rotating the figure
     about the line of centers.  The numbers labeled are $\log \xi$.  The
     $\times$ at (0,0) marks the the position of the neutron star, and the
     shadow cone corresponds to region in which the compact X-ray source is
     occulted by the companion. The coordinate axes are scaled in units of
     solar radii.}  \label{fig3}
\end{figure}

\begin{figure}
  \caption{ The apparent {\small DEM} at three different orbital phases.  The
    top curve is for phase, $\phi = 0.5$, which corresponds to the intrinsic
    {\small DEM} of the system.  The middle and bottom curves are the apparent
    {\small DEM} curves at quadrature and eclipse of the X-ray source,
    respectively.  The apparent emission measure during eclipse is 30\% of the
    total emission measure.}  \label{fig4}
\end{figure}

\begin{figure}
  \caption{ The apparent {\small DEM} curves for $\beta = 0.0$ to $1.0$ in
    steps of 0.2, averaged over the Vela X-1 eclipse phase.  All other
    parameters are shown in Table 4.}  \label{fig5}
\end{figure}

\begin{figure}
  \caption{ The 90\% and 99\% $\chi^2$ contour levels for a smooth wind model.
    Also shown are two sets of wind parameters derived using different
    techniques.}  \label{fig6}
\end{figure}

\begin{figure}
  \caption{ The model {\small DEM} that gives the best-fit to the {\small SIS}
    data.  The parameters are shown in Table 4.}  \label{fig7}
\end{figure}

\begin{figure}
  \caption{ A comparison of the data with the model generated with the {\small
    DEM} curve shown in Figure 7.  The resulting $\chi^2_r$ is 1.01 for 296
    degrees of freedom.  {\small SIS}1 data is not shown for clarity.}
    \label{fig8}
\end{figure}

\begin{figure}
  \caption{ The product of the equivalent hydrogen column density and the
    covering fraction versus the atomic number of the detected fluorescent
    lines derived from the measured intensities shown in Table 2.  The two
    points at $Z = 26$ correspond to the column densities derived from iron
    K$\alpha$ (lower point) and K$\beta$ (higher point).  A monotonic
    dependence on $Z$ is observed.}  \label{fig9}
\end{figure}

\clearpage

\begin{deluxetable}{cccr}
\tablecolumns{4}
\tablecaption{Energies and Ionization Parameters of Formation for
              Bright Recombination Lines \label{table1}}
\tablehead{
  \colhead{Ion} &
  \colhead{Line energy (keV) \tablenotemark{a}} &
  \colhead{$\xi_{\rm{form}}$} &
  \colhead{$T_{\rm{form}}$ (eV)}
  }

\startdata
 O VII   & 0.562 & 1.49 &   7 \nl
 O VIII  & 0.654 & 1.92 &  18 \nl
Ne IX    & 0.910 & 1.76 &  12 \nl
Ne X     & 1.022 & 2.30 &  97 \nl
Mg XI    & 1.337 & 1.98 &  21 \nl
Mg XII   & 1.473 & 2.58 & 123 \nl
Si XIII  & 1.847 & 2.15 &  75 \nl
Si XIV   & 2.005 & 2.73 & 145 \nl
 S XV    & 2.441 & 2.39 & 105 \nl
 S XVI   & 2.622 & 2.90 & 183 \nl
Fe XXV   & 6.667 & 3.11 & 250 \nl
Fe XXVI  & 6.966 & 3.66 & 650 \nl
\tablenotetext{a}{Centroid energy of the line complex}
\enddata

\end{deluxetable}

\clearpage

\begin{deluxetable}{cccr}
\tablecaption{Measured Fluorescent Line Energies and Intensities
              \label{table2}}
\tablehead{
  \colhead{Line Energy (keV)} &
  \colhead{Average Ionization Stage}\tablenotemark{a} &
  \colhead{Flux (photon cm$^{-2}$ s$^{-1}$)}\tablenotemark{b} &
  \colhead{$\Delta \chi^2$}\tablenotemark{c}
  }
\startdata
1.30$^{+0.02}_{-0.03}$ & Mg {\small II-VIII} &
   $(2.2^{+0.8}_{-0.6}) \times 10^{-5}$ & 15 \nl
1.77$^{+0.02}_{-0.03}$ & Si {\small I-V} &
   $(2.0^{+0.7}_{-0.5}) \times 10^{-5}$ & 21 \nl
2.32$^{+0.03}_{-0.02}$ & S {\small I-IV} &
   $(2.6^{+0.8}_{-0.7}) \times 10^{-5}$ & 33 \nl
2.99$^{+0.03}_{-0.02}$ & Ar {\small II-V} &
   $(2.3^{+0.6}_{-0.5}) \times 10^{-5}$ & 26 \nl
3.74$^{+0.05}_{-0.06}$ & Ca {\small I-VIII} &
   $(1.5^{+0.5}_{-0.5}) \times 10^{-5}$ & 11 \nl
6.45$^{+0.01}_{-0.01}$ & Fe {\small IV-V} (K$\alpha$) &
   $(2.5^{+0.2}_{-0.2}) \times 10^{-4}$ & 242 \nl
7.13$^{+0.04}_{-0.05}$ & Fe {\small II-V} (K$\beta$) &
   $(7.2^{+1.7}_{-1.6}) \times 10^{-5}$ & 31 \nl
7.52$^{+0.06}_{-0.05}$ & Ni {\small I-VIII} &
   $(2.2^{+1.5}_{-1.4}) \times 10^{-5}$ & 3 \nl
\tablecomments{All errors correspond to 90\% confidence ranges for one
               interesting parameter.}
\tablenotetext{a}{The range in ionization stages are based on the
                  uncertainties in the line energies.}
\tablenotetext{b}{The widths of the lines are fixed at $\sigma=50$ eV.}
\tablenotetext{c}{The increase in $\Delta \chi^2$ when the line is
	          excluded from the spectral model.}

\enddata
\end{deluxetable}

\clearpage

\begin{deluxetable}{lcccccc}
\tablecolumns{7}
\tablecaption{Measured Continuum Parameters
                \label{table3}}
\tablehead{
  \colhead{} & \multicolumn{3}{c}{Scattered} &
  \multicolumn{3}{c}{Direct} \\
  \cline{2-4} \cline{5-7} \\
  \colhead{Orbital Phase} & \colhead{$N_H$\tablenotemark{a}} &
  \colhead{$\Gamma$\tablenotemark{b}} & \colhead{$A$\tablenotemark{c}} &
  \colhead{$N_H$\tablenotemark{a}} &
  \colhead{$\Gamma$} & \colhead{$A$\tablenotemark{c}}
  }
 
\startdata
Pre-eclipse   &  1.4 & 1.7 & $2.4 \times 10^{-3}$
              & 58.2 & 1.7 & $7.2 \times 10^{-1}$ \nl
Eclipse       &  0.5 & 1.7 & $9.0 \times 10^{-4}$
              & 23.0 & 1.7 & $9.1 \times 10^{-3}$ \nl
Post-eclipse  &  1.9 & 1.7 & $3.1 \times 10^{-3}$
              & 18.5 & 1.7 & $5.8 \times 10^{-1}$ \nl
\tablenotetext{a}{in units of $10^{22}$ cm$^{-2}$}
\tablenotetext{b}{Fixed to the corresponding absorbed powerlaw index}
\tablenotetext{c}{photon cm$^{-2}$ s$^{-1}$ kev$^{-1}$ at 1 keV}
\enddata
\end{deluxetable}

\clearpage

\begin{deluxetable}{lcl}
\tablecaption{Adopted and Derived Parameters for the Vela X-1 System
		\label{table4}}
\tablehead{
  \colhead{Parameter} &
  \colhead{Value} &
  \colhead{Reference}
  }

\startdata
$T_{\rm{eff}}$ & 26,000 K                      & Conti 1978 \nl
$a$        & 53.4 $R_\odot$                    & van Kerkwijk et al.\ 1995 \nl
$R_\ast$   & 30.0 $R_\odot$                    & van Kerkwijk et al.\ 1995 \nl
$i$        & 74$^\circ$                        & Conti 1978 \nl
$E_c$      & 20 keV                            & White, Swank, \& Holt 1983 \nl
$E_f$      & 16 keV                            & White, Swank, \& Holt 1983 \nl
$L_x$      & $4.5 \times 10^{36}$ erg s$^{-1}$ & This paper \nl
$\beta$    & 0.79                              & This paper \nl
$v_\infty$ & 1700 km s$^{-1}$                  & Dupree et al.\ 1980 \nl
$\dot{M}$  & $2.7 \times 10^{-7} M_\odot$ yr$^{-1}$ & This paper \nl
\enddata
\end{deluxetable}

\clearpage

\centerline{\psfig{figure=f1.ps,angle=-90,height=5in}}

\centerline{\psfig{figure=f2.ps,angle=-90,height=5in}}

\centerline{\psfig{figure=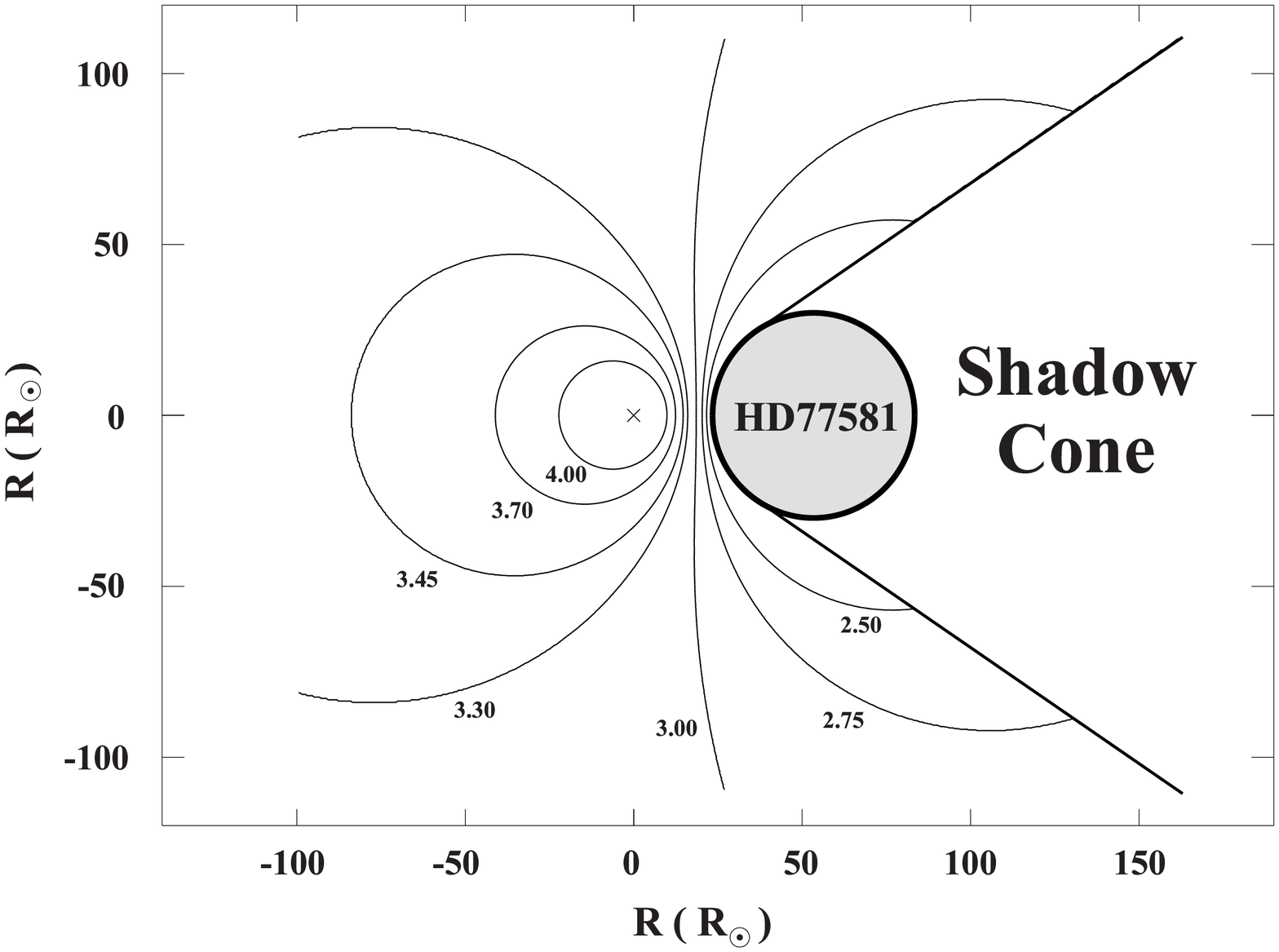,height=5in}}

\centerline{\psfig{figure=f4.ps,angle=-90,height=5in}}

\centerline{\psfig{figure=f5.ps,angle=-90,height=5in}}

\centerline{\psfig{figure=f6.ps,angle=-90,height=5in}}

\centerline{\psfig{figure=f7.ps,angle=-90,height=5in}}

\centerline{\psfig{figure=f8.ps,angle=-90,height=5in}}

\centerline{\psfig{figure=f9.ps,angle=-90,height=5in}}

\end{document}